\documentclass[submission,copyright,creativecommons]{eptcs}
\providecommand{\event}{ASQAP 2025} % Name of the event you are submitting to

\usepackage{iftex}

\ifpdf
  \usepackage{underscore}         % Only needed if you use pdflatex.
  \usepackage[T1]{fontenc}        % Recommended with pdflatex
\else
  \usepackage{breakurl}           % Not needed if you use pdflatex only.
\fi

\usepackage{comment}
\usepackage{graphicx}
\usepackage{float}
\usepackage{wasysym}
\usepackage{xcolor}

\usepackage{tabularx, booktabs}
\makeatletter
\renewcommand\paragraph{\@startsection{paragraph}{4}{\z@}%
            {-0.5ex\@plus -1ex \@minus -.25ex}%
            {.05ex \@plus .05ex}%
            {\normalfont\normalsize\bfseries}}

\title{Verification of Digital Twins using Classical \& Statistical Model Checking}
\author{Raghavendran Gunasekaran
\institute{Tilburg University}
\institute{Tilburg School of Humanities and Digital Sciences\\
Tilburg, The Netherlands}
\email{r.gunasekaran@tilburguniversity.edu}
\and
Boudewijn Haverkort\qquad\qquad
\institute{University of Twente\\
Faculty for Electrical Engineering,\\ Mathematics and Computer Science\\
Enschede, The Netherlands}
\email{\quad b.r.h.m.haverkort@utwente.nl \quad\qquad}
}  
\def\titlerunning{Verification of digital twins using classical \& statistical model checking}
\def\authorrunning{R.Gunasekaran \& B.R.H.M.Haverkort}
\begin{document}
\maketitle
%\vspace*{-6mm}
\begin{abstract}
With the increasing adoption of digital techniques, the concept of digital twin (DT) has received a widespread attention in both industry and academia. While several definitions exist for a DT, most definitions focus on the existence of a virtual entity (VE) of a real-world object or process, often comprising interconnected models which interact with each other, undergoing changes continuously owing to the synchronization with the real-world object. These interactions 
%within the VE could 
might lead to inconsistencies at execution time, due to their highly stochastic and/or time-critical nature, which may lead to undesirable 
%(and unexpected) 
behavior. In addition, the continuously varying nature of VE owing to its synchronization with the real-world object further contributes to the complexity arising from these interactions and corresponding model execution times, which could possibly affect its overall functioning at runtime. This creates a need to perform (continuous) verification of the VE, to ensure that it behaves consistently at runtime by adhering to desired properties such as deadlock freeness, functional correctness, liveness and timeliness.
%and others. Since 
Some critical properties such as deadlock freeness can only be verified using classical model checking; on the other hand, statistical model checking provides the possibility to model actual stochastic temporal behavior. We therefore propose to use both these techniques to verify the correctness and the fulfillment of desirable properties of VE. We present our observations and findings from applying these techniques on the DT of an autonomously driving truck. Results from these verification techniques suggest that this DT adheres to properties of deadlock freeness and functional correctness, but not adhering to timeliness properties. %We compare these two techniques in terms of their modeling capabilities, verifiable properties, obtained results and their performance, for verification of this DT.
\end{abstract}

\vspace*{-6mm}
\section{Introduction} \label{Intro}
\vspace*{-2.8mm}
With the advent of cloud computing, big data and artificial intelligence, the process of digitalisation takes the center stage for driving innovation. One of the key technologies for digitalisation gaining widespread attention in both industry and academia are digital twins (DTs). They have been used for several applications in industry, e.g., control, monitoring, predictive maintenance, validation, design. Since its inception in 2002, DTs have been provided with several definitions~\cite{shafto2010draft, grieves2014digital, gabor2016simulation, canedo2016industrial, tao2017digital, kritzinger2018digital}; most of these definitions focus on the existence of a virtual replica of a physical object or process (often called actual entity (AE)), which comprises several interconnected models and which synchronizes continuously with its physical counterpart. These models interact with each other to exchange data in order to effect the purpose of the DT during its execution. The myriad of interactions within a DT could cause inconsistent behavior at runtime which is undesirable~\cite{van2021models}. In addition, the continuous synchronization between the VE and AE contributes to a continuous evolution of the DT during its lifecycle, which poses additional complexity at runtime; requiring continuous verification of DTs. 

%{\em Contribution:} 
To the best of our knowledge, there exists no literature which explicitly addresses  the verification and validation (V\&V) of runtime behavior of DTs. In this paper, we propose to combine the techniques of classical model checking (CMC) and statistical model checking (SMC) to verify the run-time behavior of DTs. We apply these techniques to a DT of an autonomously driving truck 
%in a distribution center 
(which lacks complete or structured documentation) to verify the desired properties of interest using the UPPAAL toolset~\cite{behrmann2006tutorial}. The sections on the DT case study and behavioral modeling of this DT in UPPAAL has already been covered in our previous work~\cite{gunasekaran2024verification}. Due to space constraints, we omit these sections in this paper and we refer the readers to our previous work, Gunasekaran and Haverkort~\cite{gunasekaran2024verification}.
%and perform what-if explorations using the UPPAAL tool. 
Moreover, from our observations and findings, we compare and contrast SMC with CMC to guide the reader in selecting the right technique for verification of their DT. %\rg{In addition, we propose to use what-if scenario explorations with CMC in order to possibly tackle the challenge of verification of DTs as they keep varying across their lifecycle.}

\vspace*{-6mm}
%\section{Verification \& validation of DTs}
\section{Background} \label{Background}
\vspace*{-4.8mm}
The runtime behavior of a DT is the emergent behavior arising from the compositional effect of interactions between its different components. Tao et al.~\cite{tao2024advancements} and Van den Brand et al.~\cite{van2021models} point to the runtime complexity of the multitude of interactions (which could be stochastic and time-critical in nature) occurring within a DT which could lead to inconsistent or under-performing behavior, both of which are undesirable. This type of behavior can only be observed during the execution of the overall DT, i.e., when the different components in the DT interact with each other; it cannot be determined from the individual components. In order to completely avoid such undesirable behavior, verification of the runtime behavior of the DT can come to the rescue. 
%While DTs have been used for a wide range of applications, they are also used for applications of high criticality and safety. 
%Several works address DTs being used for production control purposes~\cite{kostenko2018digital,avventuroso2017networked,liu2021review} where the VE directly controls the AE by interacting with it. When used for safety-critical systems, the actuation from VE to AE needs to obey real-time constraints \cite{dalibor2022cross}. Any undesirable or unpredictable behavior in these VEs during their execution, could possibly affect the actual application (production control in this case) for which the VE is being used. This points towards a need to verify the VE. 
In addition to the complexity arising from multitude of interactions between cross-domain models, the evolution of VE as elucidated by Zhang et al.~\cite{zhang2021building} and Mertens et al.~\cite{mertens2024continuous} due to its continuous synchronization with AE and other reasons
pose further challenges.
%and other reasons as highlighted by Van den Brand et al.~\cite{van2021models} and, Gunasekaran and Haverkort~\cite{gunasekaran2024verification} 
%, which has also been elucidated by Zhang et al.~\cite{zhang2021building} and Mertens et al.~\cite{mertens2024continuous}.
%who discusses three types of evolution of VE, namely, self-to-self evolution, self-to-new evolution and VE reconstruction. 
%There are other reasons which also contribute towards this variability of VEs, such as modifications and improvements made to the PE (as a result VE adapts accordingly in order to continue staying as its faithful twin), bug fixes and others. 
This variability in VE across its  lifecycle, results in the adaptation of its models and the interactions between the models,
%, also consequently affects the interactions between these models as well leading
leading to further uncertainty issues at runtime. Thus, performing verification of VE only at its design stage does not assure its correct functioning. Consequently, the dynamic nature of the VE requires a continuous verification of its runtime behavior across its entire lifecycle.
%All these changes that the models within a VE undergo as part of its evolution might possibly affect the interactions between them, creating more uncertainty which could lead to undesirable behavior.
Existing literature of DTs focusses only on the V\&V of the visible behavior of DTs. 
%This type of validation pertains to comparing the observable behavior of AE
%the real-world object
%and VE to detect differences, %This type of validation focuses
%specifically focusing on validating the fidelity of the VE, i.e., how faithfully VE represents its physical counterpart (AE). 
Muctadir et al.'s~\cite{muctadir2024current} interview research on understanding the current practices involved in the development and maintenance of digital twins found that 13 of the 19 practitioners from both industry and academia were validating their DT by comparing the behavior between VE and AE~\cite{muctadir2024current}. Grieves et al.~\cite{grieves2017digital} discusses visual tests, performance tests, and reflectivity tests for V\&V of DTs and these tests also focus on validating the fidelity of the VE. Muñoz et al.~\cite{munoz2024measuring} also perform validations of DTs using trace alignments to check for fidelity by comparing the behavior of VE and AE. Dalibor et al.~\cite{dalibor2022cross} discuss different techniques such as simulation, testing, model checking and others for validation of DTs.  All of these validation methods focus only on validating the DTs at design time and not at run-time, nor do these discuss {\em continuous} V\&V of DTs. 

\vspace*{-6mm}
\section{Classical \& statistical model checking} \label{MC+SMC}
\vspace*{-4.5mm}
Formal verification techniques such as CMC ensure complete correctness of a system~\cite{baier2008principles}. It is a technique of modeling the behavior of a system using a model description language 
%supported by the tool used for checking the validity of certain system requirements (the so-called model checker). The 
and the requirements of the system are formally specified as properties in order to avoid ambiguity. The 
%CMC tool (so-called model checker) 
model checker performs a complete state space exploration to systematically check whether a specified property holds for (a given state in) that model~\cite{baier2008principles}. If the property is not satisfied, then most CMC tools provide a counterexample which helps in uncovering issues in the behavior which helps in refining the design of the system. 
%On the contrary, it can also be used in refining the behavioral model or the specified property. 
CMC can be applied for verifying reachability properties such as deadlock and livelock freeness, liveness properties, fairness properties and others. However, it suffers from state space explosion issues. Even for simple systems, at times the computed state spaces may become too large for realistic amount of resources.
%(computational and memory). 
%The state space explosion issue highly affects the adoption of this technique for verification of systems.
On the other hand, SMC provides probabilistic estimates based on simulations or samples of system executions~\cite{agha2018survey}. Since it relies only on samples and not complete system executions, it does not face state space explosion issues. 
%nor provides complete correctness guarantees.
While it does provide options to set the degree of statistical confidence on the probabilistic estimates, unlike CMC, it does not provide complete correctness guarantees though. Due to its reliance on samples of system execution, these techniques could possibly be applied for verification (with estimates on probability measures) for systems from a wide range of domains~\cite{basile2021analysing, legay2015statistical, david2015uppaal}. 
In addition, SMC can also be applied to systems which have complex dynamics, whose behavior is stochastic in nature~\cite{agha2018survey}. In this paper, we selected the UPPAAL tool for CMC and SMC of runtime behavior of DTs.
%, considering its multifaceted nature. 
UPPAAL SMC has a rich semantics which helps in performing simulations, probability estimations, hypothesis testing, probability comparison and value estimations.
Since the properties to be verified in a DT at runtime encompass both reachability properties (such as deadlock freeness, liveness) and temporal properties (such as occurrence of an interaction within a given time bound, estimating the delay of interactions, etc.), it appears natural to use both CMC and SMC.
%classical and statistical model checking. %DTs which are used for controlling safety critical real world entities such as a (scaled down) truck in a distribution center (cf.section~\ref{DTcasestudy}) would require to have a safe functional behavior which is devoid of errors. In such cases of DTs used for control purposes, the property of deadlock freeness becomes highly important to be verified at runtime. 

Combining and applying both these techniques will help in verifying all the desired properties of interest at runtime. Whereas literature exist on both CMC and SMC, we are not aware of any literature which combines both these techniques for verification of DTs, nor on discussions on the pros and cons of each of these techniques, given a certain case study.

\vspace*{-6mm}
\section{Combining classical and statistical model checking for verification} \label{Formal_analysis}
\vspace*{-4mm}
As detailed in~\cite{gunasekaran2024verification}, in order to verify the VE with its stochastic nature of interactions, we decided to use the exact varying timing data from histograms (actual) as it is for SMC in stochastic timed automata (STA)~\cite{bertrand2014stochastic}. For CMC, we decided to use the weighted average from these histograms (approximate) as delay values in timed automata (TA)~\cite{alur1999timed}.
This is because CMC does not provide the possibility to verify the DT with such stochastic timings. In terms of modeling in UPPAAL, both TA and STA are identical in
all aspects, except for the specification of the delay values. In order to perform verification, the next step after modeling is to formally specify the properties of interest. %which are needed to be verified in the model. 
As mentioned in section~\ref{Intro}, no documentation or formal requirements for this DT are known to us. %Thus, in order to check for the expected functionality and timeliness of the DT, we had created a list of properties.
Thus, we created a list of properties which we believe could ensure functional correctness (such as deadlock freeness, liveness and others) and timeliness of the DT. These properties are listed in Table~\ref{Table_results}. The purpose of verification of properties using both CMC and SMC is twofold: First, as mentioned in section~\ref{MC+SMC}, not all properties can be verified with just one technique and require a combination of both.
%some critical properties such as deadlock freeness and liveness can only be verified using CMC, and likewise properties pertinent to value estimations and simulations can only be performed with SMC.
Secondly, on observing the high degree of variability with the temporal behavior in interactions within VE, verification of properties performed using both approximate behavior (using CMC) and actual temporal behavior (using SMC) %(discussed in~\cite{gunasekaran2024verification})
%section~\ref{Statistical_analysis}) 
helps in understanding the differences. %(if any discrepancies in results occur).
%The verification of properties using both approximate (CMC) and actual temporal behavior (SMC) was performed to understand the difference in results (if any). Moreover, as mentioned earlier in section~\ref{MC+SMC}, not all properties can be verified using only one technique and thus, both CMC and SMC were used.
The queries for UPPAAL SMC use the formal language of MITL (Metric Interval Temporal Logic)~\cite{dokhanchi2015metric} and the queries for CMC use CTL (Computation Tree Logic)~\cite{reynolds2001axiomatization}; we refer to the UPPAAL documentation~\cite{UPPAALdoc}
%\footnote{See \url{https://docs.uppaal.org/language-reference/query-syntax/}.} 
for query syntax and semantics.
%\rg{One of the disadvantages with SMC, as mentioned in Section~\ref{MC+SMC}, is that it does not perform a complete state space exploration. Thus, in cases where stochastic time distributions are specified using a random() function in UPPAAL, queries such as deadlock freeness cannot be verified. Hence, we used CMC to verify the property of deadlock freeness and other relevant properties in VE, which is planned to be discussed in another paper.}
The system specifications, tool version and other required configurations for experimentation with CMC and SMC is the same as specified in~\cite{gunasekaran2024verification}. 
%All our experiments were performed on a an Intel(R) Core(TM) i7-10510U CPU, clocked at 1.80GHz, 2.30 GHz, which comprises 4 cores and 8 logical processors, 32GB RAM and running a 64bit Windows 10 OS. UPPAAL version 5.0.0 was used for the experiments for both CMC and SMC. For SMC, the confidence parameters, i.e., the probability of false negatives(), and the probability of false positives() and uncertainty() were all set to 5\% (Confidence Interval of 95\%).

%The next step after modeling in UPPAAL is to formally specify the properties of interest which need to be verified. All our experiments were performed on a PC with an Intel(R) Core(TM) i7-10510U CPU, clocked at 1.80GHz, 2.30 GHz, which comprises 4 cores and 8 logical processors, 32GB RAM and running a 64bit Windows 10 operating system. The UPPAAL version 5.0.0 which includes java was used for the experiments for both CMC and SMC.

\vspace*{-5mm}
\subsection{Properties and results from CMC and SMC}
\vspace*{-3mm}
We initially performed CMC to check for (safety) critical properties. 
%Since some of these properties can also be queried using SMC, we also verified these properties using SMC. Moreover, this also helps in comparing the results of the same properties from both CMC and SMC. Since SMC helps in modeling the actual behavior of system than CMC (discussed in paragraph {\em Required changes for transition from CMC to SMC} in this section), this helps in checking for discrepancies between the results obtained for actual behavior (using SMC) and approximate behavior (using CMC). 
Subsequently, SMC was performed for properties involving value estimations and simulations. SMC provides the capability to determine the worst-case response time (WCRT) of certain critical interactions and performs simulations on such critical parameters. 
%CMC was not performed on these properties since they can only be queried using SMC. 
Moreover, for the properties for both CMC and SMC, we also obtained the performance results of these properties in terms of the number of states explored, CPU time and memory consumed, which are specified in Table~\ref{Table_results}. This was done to help us compare the performance of both techniques. The queries in CTL for CMC and queries in MITL for SMC are listed in~\cite{MCQueries} (due to space constraints).

\begin{table*}[htbp] %\captionsetup{format=hang, labelsep=mysep} 
\centering
\caption{Properties and results from UPPAAL verifier for CMC and SMC} \label{Table_results}
%\vspace*{-3mm}
\begin{tabular}{ |p{0.3cm}|p{7.5cm}||p{1.2cm}|p{1.4cm}|p{1.5cm}|p{1.4cm}|}
 \hline
% and what-if exploration along with PE}   
 %\multicolumn{5}{|c|}{List of properties verified using Statistical Model Checking}\\
 %\hline
\raggedleft{ \small No.} &\centering{Properties for CMC and SMC

(Performance of CMC and SMC specified in terms of number of states explored; CPU time in ms; memory consumption in KiB)} &\centering{Results from CMC}  &\centering{Performa- -nce of CMC}  &\centering{Results from SMC} &\centering\arraybackslash{Performa- -nce of SMC}\\
 \hline
        1 & \small There is never a situation when the interactions between the two Simulink models and Python server stops, when the interactions between the two Simulink models and Unity model keeps happening.   & \small Satisfied  & \small 21869; 125; 68272 & \small Pr $\geq$ 0.963783 & \small 159426; 204; 55784  \\ \hline
        %&Pr >= 0.963783 &159426 &204 &55784
        
        %1 & There is never a situation when the interactions between the two Simulink models and Python server stops, when the interactions between the two Simulink models and Unity model keeps happening.   &Satisfied  &21869; 125; 68272 &Pr >= 0.963783 &159658; 125; 57032  \\ \hline
        
        2 & \small When the cosimulation has started, all the interactions between the two Simulink models and Unity model should eventually start happening.   & \small Satisfied  & \small 0; 0; 54360  & \small Pr $\geq$ 0.963783 & \small 0; 0; 57048  \\ \hline

        3 & \small When the cosimulation has started, all the interactions between the two Simulink models and Python server should eventually start happening.   & \small Satisfied  & \small 0; 0; 54316 & \small Pr $\geq$ 0.963783 & \small 0; 0; 57036   \\ \hline

        4 & \small When a cosimulation is started by user, it should always eventually end.  & \small Out of memory  &- &- &- \\ \hline

        5 & \small A request from Simulink path planner model eventually receives a response from Unity.   & \small Not satisfied  & \small 7011; 47; 65732 &- &-  \\ \hline

        6 & \small The DT should not have any deadlocks at runtime.   & \small Satisfied  & \small 1448411; 452578; 310868 &- &-  \\ \hline
        
         %1 & What is the probability of a situation occurring when the interactions between the two simulink models and python server will not stop while the interactions between the two simulink models and unity model continues to progress?   &Pr >= 0.999631  &Pr >= 0.999631   \\ \hline
        
        7 & \small For the constant motion of the truck in the Unity simulation engine, what is the probability of Simulink controller model sending control output to Unity repeatedly within a specific time?  & \small Not satisfied & \small 4228; 15; 63532  & \small Pr $\leq$ 0.0362167 & \small 1754; 0; 57052      \\ \hline
        
        8  & \small For continuous computation of path information, what is the probability of the Unity simulation engine sending the obstacle position information repeatedly within a specific time? & \small Not satisfied & \small 299; 0; 65192  & \small Pr $\leq$ 0.0362167 & \small 7942; 0; 57044     \\ \hline
        
        9   & \small What is the probability of the Simulink controller model receiving the path information every time before it sends the control output to Unity model?  & \small Not satisfied & \small 6; 0; 64876 & \small Pr $\leq$ 0.0362167 & \small 1200; 0; 57076   \\ \hline
        
        10    & \small What is the probability of the Simulink path planner model receiving the obstacle position information every time before it sends the path information to the Python server?  & \small Not satisfied & \small 11; 0; 64900 & \small Pr $\leq$ 0.0362167 & \small 2946; 0; 57060   \\ \hline

        11    & \small What is the probability that interactions between the Simulink controller model and Python server happens every time after the the interaction between Simulink controller model and Unity model occurs?  & \small Not satisfied & \small 67289; 343; 80364  & \small Pr $\geq$ 0.997945 & \small 158353; 140; 57116  \\ \hline

        %7    & What is the probability that the cosimulation will eventually end? &Pr >= 0.999631  &Pr >= 0.999631  \\ \hline

        12    & \small What is the maximum value of delay of sending control output by the Simulink controller model repeatedly? &-  &- & \small 1387.94 +/- 1.35 ms & \small 2000000; 2750; 57080   \\ \hline
        
        %13    &What is the maximum value of delay of sending velocity information by the Unity model repeatedly?  &- &-  &1387.38 +/- 1.33 ms & 2000000; 2344;57044    \\ \hline
        
        %14   &What is the maximum value of delay of sending the obstacle information by the Unity model repeatedly?  &- &-  &696.998 +/- 0.88 ms & 2000000; 2266;57084  \\ \hline
        
        %15  &What is the maximum value of delay of sending path information to the Simulink controller model repeatedly?  &- &- &976.239 +/- 2.03 ms & 2000000; 2032;57072    \\ \hline

        %16  &What is the maximum value of delay of sending truck position to the Simulink path planner model repeatedly? &- &-  &1009.05 +/- 2.28 ms & 2000000; 2156;57080              \\ \hline

        %17  &Simulate the occurrences of the interactions Simulink controller model output (A1,A2) and the python server sending it the path information (G). &- &- &- &     \\ \hline

        13  & \small Simulate the occurrences of the interactions Simulink pathplanner model output (E) to the python server and the Unity model sending the obstacle information (D1,D2,D3). &- &- &~\cite{gunasekaran2024verification} & \small 159918; 204; 58584  \\ \hline

        %19  &Simulate the occurrences of the interactions Simulink controller model output (F) and the Unity model sending the velocity, and truck and trailer positions (B1,B2,B3).  &- &-  \\ \hline

        %14  &Simulate the delay values of Simulink controller model output(A1,A2) and the python server sending it the path information(G).  &- &- &Figure~\ref{fig:Sim_delay_AG_1}     & 160550; 187; 58468    \\ \hline

        %21  &Simulate the delay values of Simulink pathplanner model output (E) and the Unity model sending the obstacle information (D1,D2,D3). &- &-    \\ \hline

        %22   &Simulate the delay values of Simulink controller model output( F) and the Unity model sending the velocity, and truck and trailer positions (B1,B2,B3).  &- &-  \\ \hline
 %\hline
\end{tabular}
\end{table*}
The results from Table~\ref{Table_results} are  self explanatory. All properties numbered [1-11] except for property 4 in Table~\ref{Table_results} was verified using CMC. Most of the safety critical and liveness properties in the DT were verified using CMC to ensure that the DT does not exhibit any crucial issues at runtime. The most important property to be verified was deadlock freeness and the DT satisfies this property. However, not all properties (numbered 1-6) related to functional correctness were true. Four properties (numbered 1, 2, 3, 6) which ensure functional correctness, such as ensuring that interactions between components in VE do not terminate during cosimulation, all interactions within VE eventually occurs when the cosimulation has started and the property of deadlock freeness holds true. However, property 5, which ensures that a request from the Simulink pathplanner model eventually receives a response from the Unity model does not hold. However, the liveness property, number 4, to check whether the simulation always ends eventually after it is begun by a user could not be verified
%. This was 
due to the state space explosion issue. With regard to timeliness of cosimulation of VE, none of the timeliness properties (numbered 7-11) holds true. This clearly explains that this lack of adherence to timeliness properties could be the reason why the truck crashes with the obstacle. %The reason why these timeliness properties are not true could possibly be attributed to the choice of the communication protocol used for the cosimulation, or lack of sufficient processing power or both. 
Furthermore, the lack of adherence to property 5 related to ensuring that the Simulink pathplanner model would eventually receive the response 
%(D1, D2, D3 in Figure~\ref{fig:Trucklab_VE})
containing the current obstacle information when it needs 
%(C in Figure~\ref{fig:Trucklab_VE}) 
for it to compute the correct path for avoiding that obstacle could also possibly be a contributing factor.
%for this crashing behavior of the truck. 
Queries 12 and 13 from Table~\ref{Table_results}, and other simulation and value estimation queries listed in~\cite{MCQueries}, were not executed using CMC since only SMC can provide results for these queries.
From the queries used for probability estimation (see~\cite{MCQueries}), one can observe that the queries are executed for a simulation time of 10000 time units and 100 simulations. We ran the queries for 100 simulations because after performing experiments we observed that approximately only after 75 simulations, the probability estimation queries provided probability results. We omit the results of this experiment here due to space constraints. 
From Table~\ref{Table_results}, it can be observed that the results of the probability estimation queries [1-3 and 7-11] are self-explanatory. 
%As mentioned above, properties [4-6] which are liveness and deadlock freeness properties cannot be verified using SMC. 
The WCRT obtained for property 12 is around 1387 milliseconds which is a very high delay value for sending the critical control output for seamless collision free motion of the truck in the simulation environment. 
%The results of simulation query 13 in Table~\ref{Table_results}, can be found in Figure~\ref{fig:Sim_count_DE_1}. 
The other simulation and value estimation queries used for SMC can be found in~\cite{MCQueries}. The results from simulation query 13 (omitted here due to space constraints and we refer the readers to~\cite{gunasekaran2024verification}) provides insight on the occurrences of two interactions within the DT; which when combined with property 10, can help one understand why this property 10 is not holding good~\cite{gunasekaran2024verification}.%From the queries used for probability estimation (see~\cite{MCQueries}), one can observe that the queries are executed for a simulation time of 10000 time units and 100 simulations. We ran the queries for 100 simulations because after performing experiments we observed that approximately only after 75 simulations, the probability estimation queries provided probability results. We omit the results of this experiment here due to space constraints. 

While it can be observed from Table~\ref{Table_results} that the results of properties [1-3] and [7-10] for CMC and SMC are consistent, we observed a difference for property 11.
%From Table~\ref{Table_results}, it can also be observed that the results of properties [1-3] and [7-10] for CMC and SMC are consistent. However, for property 11 we observe a difference in the verification result from CMC and SMC. 
The result for this property from CMC is "Not satisfied" and from SMC it is "Satisfied (Pr$\geq$0.997945)". This discrepancy in the result can be attributed to the fact that in SMC, the actual temporal behavior of the DT was modeled, whereas with CMC only the approximated temporal behavior of the DT was modeled (see section~\ref{Formal_analysis}). This difference in modeling 
%in SMC and CMC 
is the reason for obtaining differing response for the same property, in SMC and CMC.

\vspace*{-6.9mm}
\section{Comparison on classical and statistical model checking} \label{Discussion}
\vspace*{-4mm}
In this section we discuss differences and similarities between CMC and SMC, as experienced in our case study, in terms of modeling flexibility, performance and results. For the application of both CMC and SMC, several trade-offs exist. 
while CMC aims to ensure complete correctness of the system, SMC provides only probabilistic guarantees. However, SMC is devoid of state space explosion issues compared to the higher likelihood of this issue occurring with CMC, as observed in our case study.
In terms of modeling, SMC provides options to encode detailed timing distributions. Thus, SMC enables performing verification of systems with their actual stochastic temporal behavior, making their results more reliable in a way, though providing only probabilistic guarantees. On the other hand, though CMC provides complete correctness guarantees, such time distributions need to be approximated using statistical analysis, leading to less realistic models. The effect of this difference was clearly observed with the contrasting results obtained from verification of property 11 in our case study; while the verification of this property yielded "Satisfied" using SMC, it was "Not satisfied" using CMC. Since verification through SMC was performed on the model of the actual stochastic temporal behavior of the DT, the result obtained from SMC seems more likely to be correct.
%In terms of results, response time for model checking is much higher than SMC owing to its exhaustive state space exploration. On the hand, response time for SMC is much lower than model checking considering its use of sample executions for providing results. 
In terms of results, SMC provides more detailed results with improved visualizations~\cite{david2015uppaal} (such as timing graphs derived from simulations, probability estimates, etc.) to better understand the behavior of the system. On the other hand, CMC only provides an execution trace where a property holds, or where it does not hold. Furthermore, SMC provides the option to combine results from different techniques to help reasoning about the system behavior. On drawing parallels with CMC, one could argue that this could be done using the symbolic simulator in UPPAAL. However, using the symbolic simulator to observe the values of different variables in the behavioral model for several execution runs, is a highly cumbersome and time-consuming task which may or may not lead to these results. In this way, SMC is more advantageous than CMC in terms of combining multiple results for further analysis, and by avoiding state space explosion problems. On the other hand, critical properties requiring a complete state space exploration
%; properties 
such as deadlock-freeness and other liveness properties cannot be verified using SMC. %In terms of performance, both model checking and SMC are almost on par with each other. While memory consumption and CPU time for these techniques are almost similar, the number of states explored is lesser in model checking than SMC. As mentioned earlier, this is due to model checking performing an exhaustive state space exploration once, while SMC revisits the explored states in every simulation run.
%from theory and literature, the understanding on the performance of CMC is that it performs an exhaustive exploration which could possibly require many states to be explored and thus, requiring more memory and CPU time for processing. On the other hand, SMC uses only samples of execution and does not perform an exhaustive exploration. So, a meticulous speculation would be that SMC would perform better than CMC. However, 
By observing the performance results of both CMC and SMC in Table~\ref{Table_results}, we see that the performance of both the techniques are almost on par with each other and at times, CMC even outperforms SMC. In terms of memory consumption and CPU time, both these techniques perform very similar to each other. However, in terms of the number of states explored, CMC seems to outperform SMC, i.e., CMC explores less states than SMC in order to provide the verification results for all the properties except for property numbered 7.
%\bh{I don't understand this statement: CMC explores all states, albeit all just one; SMC explores many states, but mostly also multiple times); see the next senstence; integrate these better} \rg{I agree. I have clarified this in the text now.}
% where CMC explores lesser number of states than SMC in order to obtain the verification results. 
This could possibly be attributed to the reason that CMC explores all the states in the state space, albeit only once. On the other hand, SMC explore multiple states but, mostly revisits the explored states in every simulation run.
While each of these techniques stand out with their advantages, from our experience of applying both techniques, we advocate that combining them is most advantageous. 
%Firstly, it 
It is imperative to understand the strengths of each of these techniques in order to apply them for verification of 
%(a specific facet of) the 
systems. 
%We first made an attempt using CMC to verify these critical properties in the DT. For obtaining WCRT values and performing simulations on critical parameters in DT, 
%with minimal changes in the behavioral model,
%we applied SMC to verify other properties in the DT. Moreover, combining multiple results from SMC helped in better understanding the runtime behavior of DTs.
%While all the properties were verified with a combination of both these techniques, we weren't able to verify one of the liveness properties in the DT. 
%To aid the reader with the selection of a suitable technique for verification of runtime behavior of their DT, we quote Haverkort~\cite{haverkort1998performance}, where he mentions that the measure of interest and its required accuracy highly determines the choice of behavioral model and the corresponding verification technique to be used. 

\vspace*{-6mm}
\section{Conclusion, challenges and future work} \label{Conclusion}
\vspace*{-4mm}
%As mentioned before, to our knowledge, there currently exists no literature which explicitly focuses on V\&V of runtime behavior of DTs. 
In this paper, we propose to combine CMC and SMC for (continuous) verification of runtime behavior of DTs as they keep varying across their lifecycle. %We also performed what-if scenario explorations with CMC. 
Due to lack of proper documentation or requirements for the DT, we created a list of properties which we believe would ensure deadlock freeness, timeliness and functional correctness. We verified all but one property; owing to state-space explosion issues we could not verify a liveness property in the DT. CMC helped in verifying the critical properties such as deadlock freeness and SMC helped in determining WCRT of critical interactions within DT and performing simulations. The results from the verification helped us understand that though the DT adheres to some properties of functional correctness, it does not adhere to timeliness properties. 
%This lack of adherence to timeliness properties could be attributed to the truck colliding against the obstacle in the simulation environment. 
%We believe that the reason for lack of adherence to timeliness properties could possibly be due to lack of processing resources or the choice of communication protocol used for cosimulation or both. 
While we verified the runtime behavior of the VE against several properties using both techniques, the experience from this research activity helped us to draw comparisons between the two techniques, in terms of their modeling capability, performance and obtained results. On comparing the results, for one property we observed that SMC which helps better with modeling the actual stochastic temporal behavior of interactions within VE gave a differing result when compared to CMC, which only helps in modeling the approximate temporal behavior. %\rg{We had performed what-if scenario explorations with CMC to understand effects of speeding up interactions of components within VE.}
%Moreover, we had several learnings from SMC such as a change in a DT, in terms of addition of new interactions at runtime, does not necessarily slow down other interactions, but it may even speed up other interactions in the DT.
%Furthermore, we found that SMC can also be used for exploring design alternatives for a DT through iterative modeling and verification.
Extensive logging in DTs could possibly pose several complexities. 
%Logging which was used to uncover the runtime behavior in this case study may
%However, it can be comprehended that 
%As mentioned earlier, logs generated during execution 
%only cover the actual "observed run" of DT and not the "unobserved run" of DT which may possibly be unsafe or undesirable at times. This limitation could possibly be handled to some extent by performing what-if scenario explorations~\cite{gunasekaran2024verification}. 
Moreover, there is a possibility for gaps in the manual extraction of models for verification, from the logs generated in the DT. (Semi-) automating the model extraction process could possibly help in overcoming these gaps. We are currently exploring model learning techniques such as 
passive learning (state merging based algorithms, process mining and others), which could potentially help in automating the process of obtaining verifiable models. The conclusions on comparison between SMC and CMC is based on a single DT case study; performing a series of experiments with different DT case studies could provide further insights. We plan to extend this work by including AE and effects of its bi-directional synchronization in the analysis. 
\vspace*{-7.4mm}
\section*{Acknowledgements}
\vspace*{-4mm}
This research was funded by NWO (The Dutch national research council) under the NWO AES Perspectief program on digital twins (project P18-03/P3). 

\vspace*{-6mm}
%\nocite{*}
\bibliographystyle{eptcs}
\bibliography{generic}
\end{document}